\documentclass[a4paper,bibtotoc]{scrartcl}
\usepackage{amsmath}
\usepackage{amssymb}
\usepackage[latin9]{inputenc}
\usepackage{amsthm}
\usepackage{setspace}
\usepackage{longtable}

\begin{document}
\theoremstyle{definition}
\newtheorem{theorem}{Theorem}
\newtheorem{definition}[theorem]{Definition}
\newtheorem{lemma}[theorem]{Lemma}
\newtheorem{deflemma}[theorem]{Definition \& Lemma}
\newcommand{\dom}{\operatorname{dom}}
\newcommand{\Lev}{\operatorname{Lev}}
\newcommand{\range}{\operatorname{range}}

\author{Arno Pauly\thanks{Computer Laboratory, University of Cambridge,
  Cambridge CB3 0FD, United Kingdom} \\ \small Arno.Pauly@cl.cam.ac.uk           }
\title{Infinite Oracle Queries in Type-2 Machines (Extended Abstract)}
\maketitle

\begin{abstract}We define Oracle-Type-2-Machine capable of writing infinite oracle queries. In contrast to finite oracle queries, this extends the realm of oracle-computable functions into the discontinuous realm. Our definition is conservative; access to a computable oracle does not increase the computational power.

Other models of real hypercomputation such as \textsc{Ziegler}'s (finitely) revising computation or Type-2-Nondeterminism are shown to be special cases of  Oracle-Type-2-Machines. Our approach offers an intuitive definition of the weakest machine model capable to simulate both Type-2-Machines and BSS machines.
\end{abstract}
\section{Motivation \& Overview}
\label{introduction}
As there are several distinct notions of computability for functions on uncountable sets (primarily Type-2-Machines \cite{weihrauchd} and BSS-machines \cite{blum}), a robust framework of hypercomputation could be extremely useful to allow mutual comparisons. However, so far a variety of different concepts of real hypercomputation have been introduced and studied (\cite{ziegler2}, \cite{ziegler3}, \cite{ziegler4}, \cite{hotz}). The desirable status of a powerful unified theory of real hypercomputation has not been reached yet.

The theory of discrete hypercomputation is dominated by the concept of oracle machines, which give rise to the partial order of relative computability on the Turing degrees of problems. While a direct addition of discrete oracles to machines working on infinite sequences was used successfully to describe the relationship between computability and continuity on infinite sequences, many useful properties of oracle computation are lost in this process: Problems cannot be compared using oracle machines, since problems and oracles are no longer are exchangeable.

Recent progress on the study of Weihrauch degrees (\cite{paulyreducibilitylattice}, \cite{brattka2}) has uncovered certain similarities to the theory of Turing degrees; in fact the order theoretic properties of Weihrauch degrees are nicer, since they even form a (complete\footnote{While the continuous version of Weihrauch reducibility allows the formation of arbitrary suprema and infima, for issues of uniformity, the computable version only allows finite limits.}) distributive lattice. While it seems natural to consider Weihrauch degrees to represent degrees of incomputability, a corresponding model of hypercomputation is missing.

The major obstacle for obtaining such a model is the fact that Oracle-Type-2-Machines would need the ability to pose queries of infinite length to the oracle and process the (possibly infinite) answer. Admitting computation steps to be indexed by transfinite ordinals $\omega$, $\omega + 1$, $\omega + 2$, $\ldots$ would allow to continue computation after the query has been answered, however, even without access to an oracle this would increase the computational power\footnote{A machine computing $LPO$ (s. Definition \ref{deflpo}) could proceed as follows: In Stage 0, write a 0 in the first cell of a working tape. In stage $n$, check the $n$th input cell. If there is not a $0$, write a $1$ in the first cell of the working tape. In stage $\omega$, copy the first cell of the working tape to the output tape. Continue to output $0$.}.

In the present paper we will present a machine model that allows to pose infinite queries to oracles, without increasing the computational power beyond standard Type-2-Machines in the case of computable oracles. We will show how this model gives rise to Weihrauch degrees as associated reducibility. Several previously introduced concepts of real hypercomputation will be shown to be equivalent to oracle computation with respect to certain oracles.

\section{Foundations}
\subsection{The Model}
\label{model}
To introduce the definition of an Oracle-Type-2-Machine, we start with recalling a formal definition of Type-2-Machines. For the sake of simplicity, we consider the alphabet\footnote{In some cases, the alphabet $\mathbb{N}$ is more convenient. As standard encodings are available, we will neglect the details.} $\sum = \{0, 1\}$, one input tape, two working tapes and one output tape. A Type-2-Machine $M$ is a labelled directed graph, fulfilling the following conditions: \begin{enumerate} \item Vertices with out-degree two carry the labels $t_i$ with $i \in \{0, 1, 2\}$. We assume that the outgoing edges can be distinguished, and will refer to the first or second successor. \item Vertices with out-degree 1 carry the labels $l_i$, $r_i$ with $i \in \{0, 1, 2\}$ or $w_i^b$ with $i \in \{1, 2, 3\}$ and $b \in \{0, 1\}$.\item There must be a unique source with out-degree $1$, labelled $s$. \item The possible labels for sinks are $a$ or $r$. There are no vertices with higher out-degree.\end{enumerate}

A configuration of a Type-2-Machine $M$ is a tuple $(q, w_0, n_0, w_1, n_1, w_2, n_2, w_3, n_3)$, where $q$ is a vertex in $M$, $w_i \in \{0, 1\}^\mathbb{N}$ are infinite sequences, and $n_i \in \mathbb{N}$ natural numbers for $i \in \{0, 1, 2, 3,\}$. $w_i$ is to be interpreted as the current content of the $i$th tape, counted in the order input tape, first working tape, second working tape, output tape; $n_i$ is the current position of the reading head.

For a vertex $q$, we let $L(q)$ denote its label, and $S(q)$ its successor (or $S_0(q)$ its first and $S_1(q)$ its second successor). For a sequence $w \in \{0, 1\}^\mathbb{N}$ and a natural number $n \in \mathbb{N}$, $w[n]$ denotes the $n$th symbol in $w$. If additionally $b \in \{0, 1\}$, then $w\backslash[n=b]$ denotes the sequence which is equal to $w$ in all positions except the $n$th, which is $b$.

The standard relation $\rightarrow$ of single step transitions between configurations is defined as follows:
\begin{enumerate}
\item $(q, w_0, n_0, w_1, n_1, w_2, n_2, w_3, n_3) \rightarrow (S(q), w_0, n_0, w_1, n_1, w_2, n_2, w_3, n_3)$, if $L(q) = s$
\item $(q, w_0, n_0, w_1, n_1, w_2, n_2, w_3, n_3) \rightarrow (S_{w_i[n_i]}(q), w_0, n_0, w_1, n_1, w_2, n_2, w_3, n_3)$,\linebreak if $L(q) = t_i$
\item $(q, w_0, n_0, w_1, n_1, w_2, n_2, w_3, n_3) \rightarrow (S(q), w_0, n_0 \pm 1, w_1, n_1, w_2, n_2, w_3, n_3)$, \linebreak if $L(q) = r_0$ ($L(q) = l_0$ and $n_0 - 1 \in \mathbb{N}$)
\item $(q, w_0, n_0, w_1, n_1, w_2, n_2, w_3, n_3) \rightarrow (S(q), w_0, n_0, w_1, n_1 \pm 1, w_2, n_2, w_3, n_3)$, \linebreak if $L(q) = r_1$ ($L(q) = l_1$ and $n_1 - 1 \in \mathbb{N}$)
\item $(q, w_0, n_0, w_1, n_1, w_2, n_2, w_3, n_3) \rightarrow (S(q), w_0, n_0, w_1, n_1, w_2, n_2 \pm 1, w_3, n_3)$, \linebreak if $L(q) = r_2$ ($L(q) = l_2$ and $n_2 - 1 \in \mathbb{N}$)
\item $(q, w_0, n_0, w_1, n_1, w_2, n_2, w_3, n_3) \rightarrow (S(q), w_0, n_0, w_1\backslash[n_1=b], n_1 + 1, w_2, n_2, w_3, n_3)$, if $L(q) = w_1^b$
\item $(q, w_0, n_0, w_1, n_1, w_2, n_2, w_3, n_3) \rightarrow (S(q), w_0, n_0, w_1, n_1, w_2\backslash[n_2=b], n_2 + 1, w_3, n_3)$, if $L(q) = w_2^b$
\item $(q, w_0, n_0, w_1, n_1, w_2, n_2, w_3, n_3) \rightarrow (S(q), w_0, n_0, w_1, n_1, w_2, n_2, w_3\backslash[n_3=b], n_3 + 1)$, if $L(q) = w_3^b$
\end{enumerate}

The relation $\rightrightarrows$ of arbitrary transitions is derived as the reflexive and transitive closure of $\rightarrow$. Now there are two ways for a Type-2-Machine to have a valid output, being in a certain configuration. We will define a generalized output relation depending on a transition relation, as the details of our definition of the transitions are not relevant for the definition of the output obtained from a certain configuration.

For finite computation, $M$ yields the output $w$ starting from a configuration \linebreak $(q, w_0, n_0, w_1, n_1, w_2, n_2, w_3, n_3)$ given a transition relation $\Rightarrow$, denoted as \linebreak $(q, w_0, n_0, w_1, n_1, w_2, n_2, w_3, n_3) \rightsquigarrow_{\Rightarrow} w$, if there is a configuration \linebreak $(\hat{q}, \hat{w}_0, \hat{n}_0, \hat{w}_1, \hat{n}_1, \hat{w}_2, \hat{n}_2, w, \hat{n}_3)$ with $$(q, w_0, n_0, w_1, n_1, w_2, n_2, w_3, n_3) \Rightarrow (\hat{q}, \hat{w}_0, \hat{n}_0, \hat{w}_1, \hat{n}_1, \hat{w}_2, \hat{n}_2, w, \hat{n}_3)$$ and $L(\hat{q}) = a$\footnote{In most cases it is more convenient to consider only the first $\hat{n}_3$ symbols of $w$ as the output. In our case, however, this would just make the definition of oracle calls more complicated.}. Due to the definition, $w$ and $w_3$ will be equal except for the positions between $n_3$ and $\hat{n}_3$.

For infinite computation, a configuration $(q, w_0, n_0, w_1, n_1, w_2, n_2, w_3, n_3)$ yields output $w$ given a transition relation $\Rightarrow$, denoted as $(q, w_0, n_0, w_1, n_1, w_2, n_2, w_3, n_3) \rightsquigarrow_{\Rightarrow} w$, if for each $m \geq n_3$ there is a configuration $(q^m, w_0^m, n_0^m, w_1^m, n_1^m, w_2^m, n_2^m, w_3^m, m)$ with $(q, w_0, n_0, w_1, n_1, w_2, n_2, w_3, n_3) \Rightarrow (q^m, w_0^m, n_0^m, w_1^m, n_1^m, w_2^m, n_2^m, w_3^m, m)$ and $\lim \limits_{m \to \infty} w_3^m = w$. Note that the existence of suitable configurations always ensure the convergence of the sequence $(w_3^m)_{m \in \mathbb{N}}$.

The definition of the output of a Turing machine $M$ on input $x$ is defined as the output of $M$ from configuration $(q, x, 0, 0^\mathbb{N}, 0, 0^\mathbb{N}, 0, 0^\mathbb{N}, 0)$ with $L(q) = s$.

An Oracle-Type-2-Machine has a further possible label $?$ for vertices with out-degree $2$. However, the definition of the transition relation and the result relation are now intertwined. We use $\rightarrow_0$ to denote the transition relation for normal Type-2-Machines, that is for $\rightarrow_0$ vertices labelled $?$ effectively have the same effect has the \emph{reject}-vertices labelled $r$. We now define a sequence of transition relations and result relations \linebreak $(\rightarrow_n, \Rightarrow_n, \rightsquigarrow_{\Rightarrow_n})_{n \in \mathbb{N}}$ inductively. The oracle in use is a multi-valued function \linebreak $\mathcal{O} :\subseteq \{0, 1\}^\mathbb{N} \rightrightarrows \{0, 1\}^{\mathbb{N}}$.

$\rightarrow_n$ always includes $\rightarrow_{n - 1}$, and additionally the following transitions: $$(q, w_0, n_0, w_1, n_1, w_2, n_2, w_3, n_3) \rightarrow_n (S_0(q), w_0, n_0, w_1, n_1, y, 0, w_3, n_3)$$ if $(S_1(q), w_0, n_0, w_1, n_1, w_2, n_2, 0^\mathbb{N}, 0) \rightsquigarrow_{\Rightarrow_{n-1}} x$ and $y \in \mathcal{O}(x)$ holds. $\Rightarrow_n$ is the reflexive and transitive closure of $\rightarrow_n$.

Informally, this means that a $?$-vertex spawns a copy of the original machine, with erased output tape and with the second successor of the current vertex, retrieves the result, feeds it to the oracle, and moves the first successor vertex, with the result of the oracle query being written on the second working tape.

For each step of the inductive process above, a multi-valued function computed by the Type-2-Oracle-Machine $M$ with oracle $\mathcal{O}$ and query depth $n$ can be defined as: $$F_n(x) = \{y \mid (q, x, 0, 0^\mathbb{N}, 0, 0^\mathbb{N}, 0, 0^\mathbb{N}, 0) \rightsquigarrow_{\Rightarrow_n} y, L(q) = s\}$$If $\mathcal{O}$ is single-valued, then so will be all $F^n$. It is straight-forward to see that $F_n$ always extends $F_{n-1}$, that is $F_{n-1}(x) \subseteq F_n(x)$. If the sequence stabilizes, that is if there is an $n_0$ with $F_{n_0} = F_{n}$ for all $n \geq n_0$\footnote{Obviously, $F_{n_0} = F_{n_0 + 1}$ is already sufficient.}, we call $F_{n_0}$ the multi-valued function computed by $M$ and oracle $\mathcal{O}$.

\subsection{Basic properties}
A fundamental requirement for our definition to match the intuition of oracle computation is that an Oracle-Type-2-Machine with access to a computable oracle can compute exactly those functions computable by a normal Type-2-Machine. As Type-2-Machines are by our definition a special case of Oracle-Type-2-Machines (without $?$ being used as a label, all relations $\rightarrow_n$ are identical to $\rightarrow_0$), we certainly do not loose computational power. However, it might be possible that the mechanism used to pose the oracle questions could be abused to perform hypercomputation, even without a non-computable oracle.

To prove the contrary, we make use of the next lemma. Informally, it states that, provided a fixed query depth, there is no need to use any vertex  in different query layers.

\begin{deflemma}[Separation of Query Layers]
\label{layerseparation}
Let $M$ be an Oracle-Type-2-Machine computing $F$ with query depth $n$ and oracle $\mathcal{O}$. We define another Oracle-Type-2-Machine $\hat{M}$ computing $F$ with query depth $n$ and oracle $\mathcal{O}$, in which each vertex $v$ is labelled additionally with a natural number $N(v) \leq n$ such that the following properties are fulfilled:
\begin{enumerate}
\item $N(S(v)) = N(v)$
\item $N(S_1(v)) = N(v)$
\item $N(S_2(v)) = N(v)$ for $v = t_i$
\item $N(S_2(v)) = N(v) + 1$ for $L(v) = \ ?$
\end{enumerate}

We make $n$ identical copies of $M$ without the starting vertex $q$ with $L(q) = s$, and and number all vertices in the $i$th copy with $i$.  Starting with $i = 1$ and proceeding to $i = n - 1$ step by step, always define $S_2(v) := S_2(w)$ for all vertices $v$ with $L(v) = \ ?$ and $N(v) = i$, where $w$ is the corresponding vertex to $v$ in the $i + 1$th copy. For all vertices $v$ with $L(v) = \ ?$ and $N(v) = n$, replace the label $?$ by $r$ and delete the outgoing edges.

The resulting graph forms an Oracle-Type-2-Machine $\hat{M}$ fulfilling the desired conditions. That $\hat{M}$ indeed computes the same function with query depth $n$ as $M$ does can be checked by following the definition of the result relation.
\end{deflemma}

\begin{theorem}
Let $\mathcal{O}$ be a computable oracle and $M$ an Oracle-Type-2-Machine computing $F$ with query depth $n > 0$. Then there is another Oracle-Type-2-Machine $M'$ computing $F$ with oracle $\mathcal{O}$ and query depth $n - 1$.
\begin{proof}
We replace $M$ by $\hat{M}$ due to Definition \ref{layerseparation}. We consider all vertices $v$ with $L(v) = \ ?$ and $N(v) = n - 1$. If there is none, then no vertex can have the label $n$, and we are done. Otherwise, we show how the number of such vertices can be reduced by one, iterated application yields the result.

The construction corresponds to the one used for concatenation of Type-2-Machines. Once the chosen vertex $v$ labelled $?$ is reached, a flag is set. Whenever a vertex labelled $t_2$ occurs, it will be preceded by checking the flag: If it is not set, the normal vertex is used. Otherwise, a certain Type-2-Machine is simulated\footnote{The tapes needed for the simulated can be encoded into the first working tape without significant problems for the main computation.} until it produces the next output bit. The machine used corresponds to the concatenation of the remaining machine consisting of the vertices labelled by $n$, with an additional initial vertex connected to the former right successor of $v$, and the Type-2-Machine computing the oracle. Movement commands on the third tape (which is the second working tape) are forwarded to the second machine, as well.

For this approach to work, only a finite number of simulations might be concurrently running. For multiple calls at the same layer, each new call overrides the result from the last call (save the information that has been copied to the first working tape), which allows to abort the associated simulations. To prevent an infinite number of simulations occurring on different nesting levels, we had to restrict our considerations to finite query depth.
\end{proof}
\end{theorem}

Another desirable property of any definition of computational devices is closure under composition. Here we will show this property only for those machines possessing a distinguished query depth. For those machines where each $F_{n + 1}$ is a proper extension of $F_{n}$, weaker statements can be obtained in the same fashion.

\begin{theorem}
Let $M_i$ be an Oracle-Type-2-Machine computing the multi-valued function $F_i$ using the oracle $\mathcal{O}$ for $i \in \{0, 1\}$. Then there is an Oracle-Type-2-Machine $M$ computing the multi-valued function $F_1 \circ F_0$ using the same oracle $\mathcal{O}$.
\begin{proof}
The standard procedure to concatenate Type-2-Machines can be adapted directly to Oracle-Type-2-Machines. If the query depth $n_i$ is sufficient to guarantee stabilization for the machine $M_i$, then $M$ stabilizes at $n_0 + n_1$, as can be verified directly following the definitions.
\end{proof}
\end{theorem}

\subsection{Limiting the number of oracle queries}
While the query depth can be used to limit the power of an Oracle-Type-2-Machine, a finer distinction will prove to be useful. Instead of just limiting the nesting depth of queries, their number can be restricted. Two different concepts can be pursued, either the total number is considered, or just the number of calls made at the top level of nestings. We will primarily consider a limited number of oracle calls in cases where the query depth is limited to $1$ anyway, in which case both notions coincide.

\section{Relations of Relative Computability}
\label{reducibilities}
Similar to the several different notions of relative computability used for Turing machines and the corresponding oracle machines, one can introduce several notions of relative computability using Oracle-Type-2-Machines. It turns out that some of the most interesting ones coincide with reducibilities already suggested and studied elsewhere.

\subsection{Single Oracle Calls}
If a state labelled $?$ may occur at most once during any run of the Oracle-Type-2-Machine, we can split the Oracle-Type-2-Machine into two regular Type-2-Machines using Lemma \ref{layerseparation}: One computes the oracle query given the input, the other one computes the output, given the input and the answer to the query. Thus, the corresponding oracle computability reduction coincides with computable Weihrauch reducibility\footnote{In previous work, this reducibility has also been called Wadge reducibility $(\leq_w)$ or Type-2-Reducibility ($\leq_2$).} ($\leq_W$), defined as:

\begin{definition}
For functions $f$, $g$; $f \leq_W g$ holds, if there are computable partial functions $F$, $G$ with $f(w) = F(w, g(G(w)))$ $(w \in \dom(f))$. For multi-valued functions $A$, $B$; $A \leq_W B$ holds, if there are computable partial functions $F$, $G$ with $x \mapsto F(x, g(G(x))) \in A$ for all $g \in B$.
\end{definition}

\begin{theorem}
\label{oracleweihrauch}
A (multi-valued) function $f$ is computable with a single call to an oracle for $g$, if and only if $f \leq_W g$ holds.
\end{theorem}

The continuous variant $\leq_W$ can be obtained by granting additional access to an arbitrary finite query oracle, as this suffices to compute all continuous functions. This shows that the $\leq_W$-degrees of discontinuity can all be represented by a certain infinite oracle, and vice versa. Thus, all results known about $\leq_W$ apply to infinite oracle computation (e.g. \cite{mylatz}, \cite{mylatzb}, \cite{stein}, \cite{hertling}, \cite{paulymaster}, \cite{paulyreducibilitylattice}, \cite{weihrauchc}, \cite{brattka}, \cite{brattka2}, \cite{brattka3}, just to list some of them).

One feature that shall be pointed out is transitivity: $A \leq_W B$ and $B \leq_W C$ implies $A \leq_W C$. Thus sets like \emph{All (multi-valued) functions computable using a single query to the oracle $\mathcal{O}$} are intervals regarding $\leq_W$.

\subsection{Finitely many (independent) oracle calls with query depth $1$}
If one considers more than one call to the oracle, the query depth becomes important again. For the sake of simplicity, we will always assume the query depth to be fixed to 1 in the following.

Already for two allowed oracle calls things get more complicated: One part of the Oracle-Type-2-Machine computes the query for the first call from the input, the second part computes the second query from the input and the result of the first call, the third part takes all information available and produces the output. Thus the corresponding reducibility would ask for three computable functions $F$, $G$, $H$, so that $f(x) = F(x, g(G(x)), g(H(x, g(G(x)))))$.

In addition, the relation \emph{$f$ is computable with at most $n$ oracle calls to $g$} is not transitive anymore, diminishing its appeal for further consideration. Instead, we will consider three versions of relative computability which each contain a restriction to an unspecified but finite number of oracle calls.

Provided that the different oracle calls to not depend on each other, the cartesian product of (multi-valued) functions can be employed. The following theorem will provide an important result enabling the use of cartesian products together with Weihrauch reducibility.

\begin{definition}
For functions $f: X_1 \to Y_1$, $g: X_2 \to Y_2$, define $\langle f, g\rangle: (X_1 \times X_2) \to (Y_1, Y_2)$ by $\langle f, g\rangle(x_1, x_2) = (f(x_1), g(x_2))$. For multi-valued functions $F$, $G$, define $\langle F, G\rangle$ as the set $\{\langle f, g\rangle \mid f \in F, g \in G\}$. Define $\langle f \rangle^n$ and $\langle F \rangle^n$ by iteration.
\end{definition}

\begin{theorem}
\label{cartesianweihrauch}
$f \leq_W g$ implies $\langle f \rangle^n \leq_W \langle g \rangle^n$ for each $n \in \mathbb{N}$.
\begin{proof}
If $f(x) = F(x, g(G(x)))$ holds, then also $$\langle f \rangle^n(\bar{x}) = \langle F \rangle^n(\bar{x}, \langle g \rangle^n(\langle G \rangle^n(\bar{x})))$$ is true. As computability of a function $F$ implies computability of $\langle F \rangle^n$, this completes the proof.
\end{proof}
\end{theorem}

Instead of introducing a formal definition of independent oracle calls, we will consider a class of oracles for which independence is not necessary to arrive at a succinct notion of relative computability. If the oracle has only a finite number of possible answers, then independence can be obtained by an exponential increase in the number of queries: Replace the second query by several queries, one for each possible answer to the first query, and so on.

There are three different definitions of \emph{relatively computable using only finitely many oracle calls}. The number of oracle calls could be bounded by a constant independent of the actual input, they could be bounded by a computable functions defined on the input, or unbounded, but guaranteed to be finite. The first relation is:

\begin{definition}
Let $f \leq_W^{bc} g$ holds, if there is an $n \in \mathbb{N}$, so that $f \leq_W \langle g \rangle^n$ holds. $A \leq_W^{bc} B$ holds, if there is an $n \in \mathbb{N}$, so that $A \leq_W \langle B \rangle^n$ holds.
\end{definition}

\begin{theorem}
A (multi-valued) function $f$ is computable with a fixed finite number of oracle calls to $g$ with $|\range(g)| < \mathbb{N}$, if and only if $f \leq_W^{bc} g$ holds.
\end{theorem}

\begin{theorem}
$\leq_W^{bc}$ is transitive.
\begin{proof}
We will prove the claim just for functions, the proof for multi-valued functions proceed analogously. Assume $f \leq_W \langle g \rangle^n$ and $g \leq_W \langle h \rangle^m$. Application of Theorem \ref{cartesianweihrauch} yields $\langle g \rangle^n \leq_W \langle \langle h \rangle^m \rangle^n$. Trivial consideration is enough to see $\langle \langle h \rangle^m \rangle^n \equiv_W \langle h \rangle^{nm}$, thus we have $\langle g \rangle^n \leq_W \langle h \rangle^{nm}$. Transitivity of $\leq_W$ is used to obtain $f \leq_W \langle h \rangle^{nm}$, which implies $f \leq_W^{bc} h$.
\end{proof}
\end{theorem}

The relation of relative computability where the number of oracle calls is bounded by a computable function defined on the input was suggested in \cite[Subsection 6.1]{paulyreducibilitylattice} as $\leq_{ct}$, for the sake of consistency we will call it $\leq_W^{bf}$ here. For defining it we will need the supremum for $\leq_W$, which coincides with the coproduct of functions:

\begin{definition}
\label{defsuprema2}
Let $(f_i)_{i \in \mathbb{N}}$ be a countable family of functions. Define $\lceil f_i \rceil_{i \in \mathbb{N}}$ through $\lceil f_i \rceil_{i \in N}(ix) = if(x)$. For a countable family $(F_i)_{i \in \mathbb{N}}$ of multi-valued functions, define $\lceil F_i \rceil_{i \in \mathbb{N}}$ through: $$\lceil F_i \rceil_{i \in \mathbb{N}} = \{\lceil f_i \rceil_{i \in \mathbb{N}} \mid \forall i \in \mathbb{N} \ f_i \in F_i\}$$
\end{definition}

\begin{definition}
Let $f \leq_W^{bf} g$ hold, if $f \leq_W \lceil \langle g \rangle^n \rceil_{n \in \mathbb{N}}$ holds. Let $A \leq_W^{bf} B$ hold, if $A \leq_W \lceil \langle B \rangle^n \rceil_{n \in \mathbb{N}}$ holds.
\end{definition}

\begin{theorem}
A (multi-valued) function $f$ is computable with a finite number of oracle calls bounded by a computable function to $g$ with $|\range(g)| < \mathbb{N}$, if and only if $f \leq_W^{bf} g$ holds.
\end{theorem}

\begin{theorem}
$\leq_W^{bf}$ is transitive.
\begin{proof}
Again, the proof will be done only for functions, for multi-valued functions one can proceed analogously. Assume $f \leq_W^{bf} g$ and $g \leq_W^{bf} h$. By definition, this means $f \leq_W \lceil \langle g \rangle^n \rceil_{n \in \mathbb{N}}$ and $g \leq_W \lceil \langle h \rangle^m \rceil_{m \in \mathbb{N}}$. Observe the distributivity law \cite[Theorem 6.2]{paulyreducibilitylattice}:
$$\langle f, \lceil g_i \rceil_{i \in \mathbb{N}}\rangle \equiv_2 \lceil \langle f , g_i \rangle \rceil_{i \in I}$$
Together with Theorem \ref{cartesianweihrauch} we thus have:
$$\langle g \rangle^n \leq_W \langle \lceil \langle h \rangle^m \rceil_{m \in \mathbb{N}} \rangle^n \equiv_2 \lceil \langle h \rangle^m \rceil_{m \in \mathbb{N}}$$
As this holds for all $n \in \mathbb{N}$, the property of $\lceil \ \rceil$ being the supremum in the partial order $\leq_W$ yields $\lceil \langle g \rangle^n \rceil_{n \in \mathbb{N}} \leq_W  \lceil \langle h \rangle^m \rceil_{m \in \mathbb{N}}$. Transitivity of $\leq_W$ now completes the proof.
\end{proof}
\end{theorem}

Also the third version of \emph{relatively computable with finitely many oracle calls} can be expressed using $\leq_W$ and a certain construction derived from the parallelization introduced in \cite[Section 4]{brattka2}. We will start with defining the parallelization of a (multi-valued) function defined on $\mathbb{N}^\mathbb{N}$. For that, we fix a homeomorphism $\lambda : (\mathbb{N}^\mathbb{N})^\mathbb{N} \to \mathbb{N}^\mathbb{N}$.

\begin{definition}
Given a multi-valued function $F : \subseteq \mathbb{N}^\mathbb{N} \to \mathbb{N}^\mathbb{N}$, define $\bar{F} : \subseteq (\mathbb{N}^\mathbb{N})^\mathbb{N} \to (\mathbb{N}^\mathbb{N})^\mathbb{N}$ through $\bar{F}(\prod \limits_{n \in \mathbb{N}} x_n) = \prod \limits_{n \in \mathbb{N}} F(x_n)$. Then define $\hat{F} : \subseteq \mathbb{N}^\mathbb{N} \to \mathbb{N}^\mathbb{N}$ via $\hat{F} = \lambda \circ \bar{F} \circ \lambda^{-1}$.
\end{definition}

The variant we need is obtained by prerestricting $\lambda$ to the set $\{w \in (\mathbb{N}^\mathbb{N})^\mathbb{N} \mid |\{i \in \mathbb{N} \mid w(i) \neq 0^\mathbb{N}\}| < \infty\}$. The restriction shall be denoted $\lambda_{<\infty}$. Then we can continue to define:

\begin{definition}
Given a multi-valued function $F : \subseteq \mathbb{N}^\mathbb{N} \to \mathbb{N}^\mathbb{N}$, define $\hat{F}_{<\infty} : \subseteq \mathbb{N}^\mathbb{N} \to \mathbb{N}^\mathbb{N}$ via $\hat{F}_{<\infty} = \lambda \circ \bar{F} \circ (\lambda_{<\infty})^{-1}$.
\end{definition}

To prepare for the proof of the transitivity of the corresponding reducibility relation, we show that $\hat{.}_{<\infty}$ is a closure operator regarding $\leq_W$, similar to \cite[Proposition 4.2]{brattka2}. The statements hold both for functions and multi-valued functions.
\begin{theorem}
\label{closurehatfinite}
\begin{enumerate}
\item $f \leq_W \hat{f}_{<\infty}$
\item $f \leq_W g$ implies $\hat{f}_{<\infty} \leq_W \hat{g}_{<\infty}$.
\item $\widehat{\hat{f}_{<\infty}}_{<\infty} \leq_W \hat{f}_{<\infty}$
\begin{proof}
The proof is exactly analogous to the proof of \cite[Proposition 4.2]{brattka2}.
\end{proof}
\end{enumerate}
\end{theorem}

\begin{definition}
Let $f \leq_W^f g$ hold, if $f \leq_W \hat{g}_{<\infty}$ holds. Let $A \leq_W^f B$ hold, if $A \leq_W \hat{B}_{<\infty}$ holds.
\end{definition}

\begin{theorem}
A (multi-valued) function $f$ is computable with any finite number of oracle calls to $g$ with $|\range(g)| < \mathbb{N}$, if and only if $f \leq_W^{f} g$ holds.
\end{theorem}

\begin{theorem}
$\leq_W^f$ is transitive.
\begin{proof}
Assume $A \leq_W^f B$ and $B \leq_W^f C$. Then we have $B \leq_W \hat{C}_{<\infty}$. Application of Theorem \ref{closurehatfinite} 2. and 3. yields $\hat{B}_{<\infty} \leq_W \hat{C}_{<\infty}$, by transitivity of $\leq_W$ one can obtain $A \leq_W \hat{C}_{<\infty}$, which by definition is $A \leq_W^f C$.
\end{proof}
\end{theorem}

\subsection{Infinitely many oracle calls with fixed query depth}
Infinitely many oracle calls to a function with finite range and query depth $1$ yields the relation $\leq_{\hat{W}}$ from \cite[Definition 4.3]{brattka2}. For its properties, we refer to \cite{brattka2}.

\begin{definition}
Let $f \leq_{\hat{W}} g$ hold, if $f \leq_W \hat{g}$ holds. Let $A \leq_{\hat{W}} B$ hold, if $A \leq_W \hat{B}$ holds.
\end{definition}

\begin{theorem}
A (multi-valued) function $f$ is computable with infinitely many oracle calls with nesting depth $1$ to $g$ with $|\range(g)| < \mathbb{N}$, if and only if $f \leq_{\hat{W}} g$ holds.
\end{theorem}
\section{Using $LPO$ as oracle}
The omniscience principle $LPO$ and its equivalence class have received a lot of attention in the literature, partly motivated by the fact that $LPO$ is the least discontinuous function defined on a separable space. In this section, we will explore the power of oracle access to $LPO$, applying the different restrictions introduced so far.

\begin{definition}
\label{deflpo}
Define $LPO : \{0, 1\}^\mathbb{N} \to \{0, 1\}^\mathbb{N}$ by $LPO(0^\mathbb{N}) = 0^\mathbb{N}$ and $LPO(w) = 10^\mathbb{N}$ for $w \neq 0^\mathbb{N}$.
\end{definition}

As the range of $LPO$ is finite, the relations for multiple oracle queries introduced in Section \ref{reducibilities} can be used here.
\subsection{Classical oracle computation}
As we can use a Type-2-Machine to compute functions $f: \mathbb{N} \to \mathbb{N}$ in exactly the same way as classical Turing machines, it is an interesting question which power access to uncountable oracles such as $LPO$ provides. It is easy to see that the halting problem $\emptyset'$ can be solved with a single query to $LPO$: Start the oracle call. Simulate the machine given as input, printing $0$ for each step it does not halt. If it halts, print 1, and continue to print 0s. Then the oracle returns $0^\mathbb{N}$, if the machine halts, and $10^\mathbb{N}$ otherwise.

Now assume a function $f: \mathbb{N} \to \mathbb{N}$ with $f \leq_W LPO$. In the process of writing the oracle call, an oracle machine computes a function $G: \mathbb{N} \to \{0, 1\}^\mathbb{N}$. This function $G$ can be modified to yield a computable partial function $G' :\subseteq \mathbb{N} \to \mathbb{N}$, that halts if and only if $G$ writes a $1$. Thus, the set $G^{-1}(0^\mathbb{N}) \subseteq \mathbb{N}$ is co-recursively enumerable. This implies the existence of a computable function $H$ which $n \in G^{-1}(0^\mathbb{N})$ if and only if $H(n) \notin \emptyset'$. This shows that $f$ can be computed by an oracle machine that makes one call to $\emptyset'$.

As all functions computable with oracle access to $\emptyset'$ are still continuous, $LPO$ is not computable w.r.t. $\emptyset'$. Therefore, at least for the special case of exactly one oracle call to $LPO$, we arrived at an extension of the classical degrees of oracle computability that, restricted to the classical case, coincides with the original definition.

\subsection{Fixed finite number of queries to $LPO$}
The functions $\langle LPO \rangle^n$ that arise here are identical to the functions $LPO_n$ introduced in \cite{weihrauchc}. As demonstrated in \cite{hertling}, $LPO_n$ is complete for the set of functions with Level less or equal than $n + 1$. We will consider the relationship between finitely many oracle calls to $LPO$ and the Level of a function in further detail.

If the number of oracle queries the machine can make is fixed in advance to $n$, then we can split the Oracle-Type-2-Machine with $n$ queries into an Oracle-Type-2-Machine with $n - 1$ queries and a Type-2-Machine, the latter computing the first oracle query from the input, the former computing the output given the input and the answer to the first query. In the case of $LPO$ being the oracle, this corresponds to the $\Omega^n$-continuous functions studied in \cite{mylatz}.

The process described above to replace oracle calls by independent oracle calls can be used to derive an exponential upper bound for the level of a function. However, we will used the decomposition described in the last paragraph, recycling a related proof from \cite{paulymaster}.

\begin{theorem}
\label{ncallslevel2n}
If $f$ can be computed by an Oracle-Type-2-Machine making not more than $n$ calls to an oracle for $LPO$, then $\Lev(f) \leq 2^n$.
\begin{proof}
In the case $n = 1$ we have $f \leq_2 LPO$, together with results from \cite{hertling} the claim follows. For the induction step, assume that the (multi-valued) function computed by the machine using $n - 1$ queries is $F$. We have $\mathcal{L}_{2i}(f) \subseteq \{x \mid (x, 0^\mathbb{N}) \in \mathcal{L}_{i}(F) \vee (x, 10^\mathbb{N}) \in \mathcal{L}_{i+1}(F)\}$ and $\mathcal{L}_{2i + 1}(f) \subseteq \{x \mid (x, 10^\mathbb{N}) \in \mathcal{L}_i(F) \vee (x, 0^\mathbb{N}) \in \mathcal{L}_{i + 1}(F)\}$, yielding $\Lev(f) = 2\Lev(F) = 22^{n-1}=2^n$.
\end{proof}
\end{theorem}

\subsection{Bounded finite number of queries to $LPO$}
Making a number of oracle calls to $LPO$ that is bounded by a computable function is, as explained above, equivalent to a single oracle call to $\lceil \langle LPO \rangle^n \rceil_{n \in \mathbb{N}}$. A more natural complete problem for this class is finding the minimal number in an unsorted infinite sequence of natural numbers, see \cite{paulymaster}.

\subsection{Any finite number of queries}
If the Oracle-Type-2-Machine can make any finite number of oracle queries to $LPO$, one derives an model equivalent to finitely revising computation presented in \cite{ziegler3}. The same functions are computable by a single oracle query to \textsc{Max}.

\begin{definition}
Define the partial function $MAX :\subseteq \mathbb{N}^\mathbb{N} \to \mathbb{N}$ through $MAX(w) = \max \{w(i) \mid i \in \mathbb{N}\}$.
\end{definition}

\begin{theorem}
\label{maxleqlpo}
A single oracle query to \textsc{Max} is reducible to any finite number of oracle queries to $LPO$.
\begin{proof}
For any natural number $n$ and input sequence $w$, let $w^n$ be the sequence defined by $w^n(i) = \begin{cases} 0 & w(i) \leq n \\ 1 & \textnormal{else}\end{cases}$. A machine calls the oracle $LPO$ on $w^n$ for each $n$, until the first cell of the oracle answer contains $0$ for the first time. Then $n$ is the correct output for \textsc{Max}. As long as $w$ was a valid input for \textsc{Max}, this happens in a finite number of steps.
\end{proof}
\end{theorem}

\begin{theorem}
\label{lpoleqrev}
Finitely revising computation can simulate any finite number of oracle calls to $LPO$.
\begin{proof}
Start with simulating the Oracle-Type-2-Machine. Whenever an oracle call is encountered, continue to simulate the main computational thread of the oracle machine assuming that the oracle answered $0^\mathbb{N}$. In parallel, compute the oracle query. If during the computation of any of the finitely many oracle queries another symbol than $0$ results, abort the output written so far, return to the moment in which the computation of the respective query was started, and restart from there, using $10^\mathbb{N}$ as the answer from the oracle now.
\end{proof}
\end{theorem}

\begin{theorem}
\label{revleqmax}
Finitely revising computation can be simulated by a single oracle call to \textsc{Max}.
\begin{proof}
We only need to show that the translation from $\hat{\iota}$ to $\iota$ can be computed by such an oracle machine. Compute the oracle query by reading the input and printing the highest index of a revising mark found sofar. Once the answer $n$ is obtained from the oracle, discard the first $n$ symbols from the input and output the rest.
\end{proof}
\end{theorem}

This shows that the degree of discontinuity of revising computation is the least discontinuous but discontinuous one that is closed under composition of functions. Since the composition of functions computable by BSS machines (\cite{blum}) is computable by a BSS machine, the corresponding degree of discontinuity must contain the one considered here. The other inclusion holds as well, replicating a result from \cite{ziegler4}:

\begin{theorem}
An Oracle-Type-2-Machine making a finite number of oracle calls to $LPO$ can simulate a BSS machine.
\begin{proof}
An ordinary Type-2-Machine can simulate every computation step of a BSS machine except tests on equality. Testing two real numbers for equality is equivalent to $LPO$, so a corresponding oracle call allows an Oracle-Type-2-Machine to simulate all steps of a BSS machine.
\end{proof}
\end{theorem}

If one searches for a model of computation incorporating both the capabilities of BSS machines and Type-2-Machines, without introducing unnecessary additional power, one arrives at an Oracle-Type-2-Machine making a finite number of oracle calls to $LPO$, making this a very promising machine model for the study of algorithms on the real numbers.

\begin{theorem}
The set of functions computable with finitely many oracle calls to $LPO$ is the smallest set closed under composition and products containing the Type-2-computable and the BSS-computable functions.
\begin{proof}
Due to Theorems \ref{maxleqlpo}, \ref{lpoleqrev}, \ref{revleqmax} the said set of functions is the set of functions computable with a single oracle call to \textsc{Max}. Due to Theorem \ref{oracleweihrauch}, any function in this set is of the form $x \mapsto F(x, MAX(G(x)))$, where $F$ and $G$ are computable by a Type-2-Machine. As the function $MAX$ is computable by a BSS-machine, the set is minimal.
\end{proof}
\end{theorem}

\subsection{Infinitely many oracle queries to $LPO$}
The parallelization $\widehat{LPO}$ was studied as $C$ in \cite{stein}. There are a wide variety of problems that turned out to be equivalent to $C$, we refer to \cite{brattka2} for a contemporary overview. An equivalent model of hyper-computation is the $\alpha'$-computability from \cite{ziegler2}, \cite{ziegler3}.

A higher nesting depth, fixed to $n$, corresponds to the standard generalizations of the notions discussed above: A complete function is $C^n$ (obtained as $n$-times the concatenation of $C$) and the model of hyper-computation is $\alpha^n$-computability. As shown in \cite{brattka}, the corresponding degree of discontinuity is the set of $\sum_n$-measurable functions.

\section{Other Models of Hypercomputation}
So far we have discussed how (finitely) revising computation can be expressed as use of oracle calls to $LPO$. Other models of hypercomputation are expressible in our framework, as well. We start with Type-2-Nondeterminism as introduced by \textsc{Ziegler}.

\begin{definition}The problem \textsc{UnProject} takes a name of a Type-2-Machine $M$ and an infinite sequence $x$, and asks for an infinite sequence $y$, so that $M$ accepts $\langle x, y \rangle$.
\end{definition}

\begin{theorem}
A single oracle call to \textsc{UnProject} is equivalent to Type-2-Nondeterminism.
\begin{proof}
\textsc{UnProject} can easily be solved by a nondeterministic Type-2-Machine: Guess $y$, simulate $M$ on input $\langle x, y\rangle$. If $M$ rejects, abort the computation. If $M$ accepts, copy $y$ on the output tape.

For the other direction, the nondeterministic Type-2-Machine can be split in two parts: The first part verifies the guess, and is used as input for \textsc{UnProject} together with the actual input string. The second part uses the guessed sequence (or, alternatively, the output of \textsc{UnProject}) and the input to compute the output.
\end{proof}
\end{theorem}

\begin{theorem}
$\widehat{\textsc{UnProject}}_{<\infty} \equiv_W \textsc{UnProject}$.
\begin{proof}
One direction is trivial. For the other direction, given a finite collection of $n$ Type-2-Machines a Type-2-Machine with $n$ input tapes can be constructed that accepts, if the $i$th Type-2-Machine accepts the input on the $i$th input tape for all $i \leq n$. This machine is used as input for \textsc{UnProject} together with the product of the $n$ infinite sequences.
\end{proof}
\end{theorem}

Whether $\textsc{UnProject}$ is even equivalent to $\widehat{\textsc{UnProject}}$ is left open. An answer to this question would shed a light on the robustness of Type-2-Nondeterminism. 

The results obtained so far definitely show that Type-2-Nondeterminism can be equivalently expressed in deterministic terms. Following the parallels drawn by \textsc{Ziegler} between nondeterministic Buechi-automata and nondeterministic Type-2-Machines in claiming that nondeterminism might be the more appropriate choice for infinite computation, we refer to deterministic parity automata\footnote{A language is expressible by a nondeterministic Buechi automaton, if and only if it can be expressed by a parity automaton.} and deterministic Oracle-Type-2-Machines with finite oracle access to \textsc{UnProject}; pointing out that nondeterminism can be avoided in both cases.

An example for a problem not solvable by any analytical machine is the stability of a dynamical system (\cite{hotz}). The task of constructing an Oracle-Type-2-Machine capable of solving it is trivial: Just admit a single oracle call to the problem itself. Potential further research would consider which problems are reducible to it, whether more access to the same oracle increases the computational power, and so on.

\section{Applications}
\subsection{Arithmetic Circuits}
An application for Oracle-Type-2-Machines outside the usual realm of the Type-2-Theory of Computability are given by Arithmetic Circuits as defined in \cite{pratthartmann}. Instead of the usual gates used in Arithmetic Circuits, we will replace the multiplication gate by the two following: A continuous multiplication gate: $$A \times B := \{n * m \mid n \in A \ m \in B\} \cup \{0\}$$ and a (discontinuous) test gate: $$T(A) = \begin{cases} \emptyset & A = \emptyset \\ \{0\} & \textnormal{otherwise}\end{cases}$$ Usual multiplication can be expressed by the two new gates, and both of the new gates can be expressed by the standard gates used in Arithmetic Circuits, so our modified circuits can define exactly the functions normal Arithmetic Circuits can define.

Any Arithmetic Circuit using $n$ test gates can obviously be simulated by an Oracle-Type-2-Machine making $n$ calls to an oracle for $LPO$. An application of Theorem \ref{ncallslevel2n} yields the fact that functions definable by Arithmetic Circuits always have finite level. This result directly implies many of the results presented in \cite{pratthartmann}, others follow from the observation that the look-ahead\footnote{Defining complexity measures such as look-ahead for Oracle-Type-2-Machines probably should be one of the next steps taken to further the understanding of these machines.} needed to simulate an Arithmetic Circuit is bounded by $l(n) = n$.

\subsection{The degree of discontinuity of Nash equilibria}
While the language needed to state results regarding the degree of incomputability or discontinuity of problem was present for almost two decades in the form of Weihrauch-reducibility, the concept of Oracle-Type-2-Machines allows new proof styles suitable to arrive at new results. An example of such work is \cite{paulydiscontinuousnashequilibria}, where Oracle-Type-2-Machines are used to study the degree of discontinuity shared by multiple robust divisions, solving systems of linear inequalities and finding Nash and correlated equilibria in (zerosum) games.

\end{document}